\documentclass[sigconf]{acmart}

\AtBeginDocument{%
  }

\acmDOI{}

\acmConference[Preprint]{Preprint}{2023}{HKU}
\acmPrice{}
\acmISBN{}

\usepackage{amsmath,amsfonts,bm}

\def\eqref#1{equation~\ref{#1}}
\def\1{\bm{1}}

\def\vh{{\bm{h}}}

\def\vk{{\bm{k}}}

\def\vp{{\bm{p}}}

\def\vx{{\bm{x}}}
\def\vy{{\bm{y}}}

\DeclareMathAlphabet{\mathsfit}{\encodingdefault}{\sfdefault}{m}{sl}
\SetMathAlphabet{\mathsfit}{bold}{\encodingdefault}{\sfdefault}{bx}{n}

\def\gD{{\mathcal{D}}}

\def\gP{{\mathcal{P}}}

\def\gX{{\mathcal{X}}}

\usepackage{url}
\usepackage{hyperref}
\usepackage{booktabs}
\usepackage{nicefrac}
\usepackage{xcolor}
\usepackage{graphicx}
\usepackage{cleveref}
\usepackage{wrapfig}
\usepackage{subcaption}
\usepackage{dirtytalk}
\usepackage{textcomp}
\usepackage{listings}
\usepackage{pythonhighlight}
\usepackage{amsmath}
\usepackage{amsfonts}
\usepackage{makecell}
\usepackage{multicol}
\usepackage{multirow}
\usepackage{makecell}
\usepackage{bbding}
\usepackage{colortbl}
\usepackage{pifont}
\usepackage{mathrsfs}
\usepackage{bbm}
\usepackage{bm}
\usepackage{enumitem}
\usepackage{xspace}
\RequirePackage{algorithm}
\RequirePackage{algorithmic}

\newcommand{\benchmark}{\texttt{FinBench}\xspace}
\newcommand{\method}{\texttt{FinPT}\xspace}

\begin{document}

% \title{The Name of the Title Is Hope}
\title{\method: Financial Risk Prediction with Profile Tuning on Pretrained Foundation Models}

% \author{Anonymous ICAIF 2023 submission}

\author{
  Yuwei Yin\textsuperscript{\rm 1}, 
  Yazheng Yang\textsuperscript{\rm 1}, 
  Jian Yang\textsuperscript{\rm 2}, 
  Qi Liu\textsuperscript{\rm 1}
  \\
  \textsuperscript{\rm 1} Department of Computer Science, University of Hong Kong; \textsuperscript{\rm 2} DAMO Academy, Alibaba Group\\
  \{ywyin, liuqi\}@cs.hku.hk; yangyazh@connect.hku.hk; yj411294@alibaba-inc.com
}

\renewcommand{\shortauthors}{Yin et al.}

\begin{abstract}
Financial risk prediction plays a crucial role in the financial sector. Machine learning methods have been widely applied for automatically detecting potential risks and thus saving the cost of labor.
However, the development in this field is lagging behind in recent years by the following two facts: 1) the algorithms used are somewhat outdated, especially in the context of the fast advance of generative AI and large language models (LLMs); 2) the lack of a unified and open-sourced financial benchmark has impeded the related research for years.
To tackle these issues, we propose \method and \benchmark: the former is a novel approach for financial risk prediction that conduct Profile Tuning on large pretrained foundation models, and the latter is a set of high-quality datasets on financial risks such as default, fraud, and churn.
In \method, we fill the financial tabular data into the pre-defined instruction template, obtain natural-language customer profiles by prompting LLMs, and fine-tune large foundation models with the profile text to make predictions.
We demonstrate the effectiveness of the proposed \method by experimenting with a range of representative strong baselines on \benchmark. The analytical studies further deepen the understanding of LLMs for financial risk prediction.\footnote{The code and data are released on \url{https://github.com/YuweiYin/FinPT}}
\end{abstract}



% \keywords{datasets, neural networks, gaze detection, text tagging}
% , Large Language Models



\maketitle

%%%%%%%%%%%%%%%%%%%% Section %%%%%%%%%%%%%%%%%%%%
\section{Introduction}
\label{sec:intro}

The application of machine learning methods has greatly contributed to the field of financial risk prediction~\cite{dixon2020ml_in_fin,goodell2021ai_ml_in_fin,nazareth2023fin_app_ml}. By leveraging these techniques, financial institutions are able to better assess the financial risks associated with their customers and make more informed decisions. The automation of this process has also reduced the potential for human error, ultimately leading to greater efficiency and accuracy in the evaluation of financial risk.

\begin{figure*}[ht]
  \centering
  \includegraphics[width=0.75\textwidth]{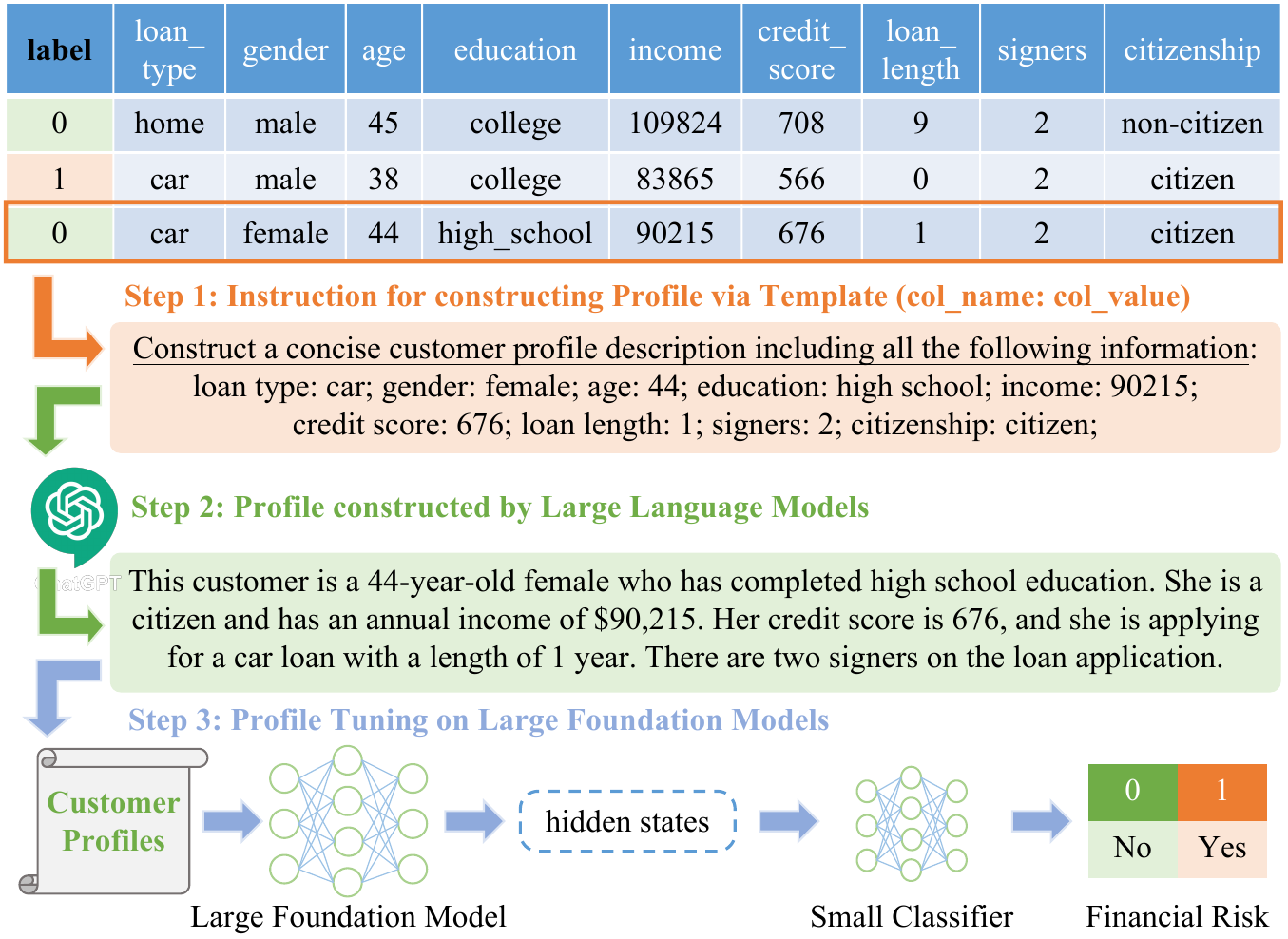}
  \caption{\textbf{The overview of \method.} \textcolor{orange}{Step 1}: fill the pre-defined instruction template with the column names and values in the table. \textcolor{teal}{Step 2}: input the instruction to large language models, such as ChatGPT and GPT-4~\cite{openai2023gpt4}, to construct customer profiles in a fluent and coherent manner. \textcolor{cyan}{Step 3}: use the natural-language profiles to fine-tune large foundation models, such as GPT~\cite{radford2019gpt2,brown2020lgpt3} and LLaMA~\cite{touvron2023llama}, with a small classifier--usually a feedforward network--to make the financial risk prediction.}
  \label{fig:finpt_pipeline}
  \Description{The overview of \method.}
\end{figure*}

% Contribution 1
Through analysis of extensive financial datasets, it has been observed that financial risk tasks can be simplified to profile classification tasks. These tasks involve evaluating customer profiles to determine if they are likely to violate financial rules. The customer profile typically includes personal information such as age, gender, education and work history, social status, and previous financial records.
Based on the fact that various financial tables share similar column names, we propose to unify the information across different tables and conduct large-scale model training. Also motivated by the blooming development and excellent ability of large language models (LLMs)~\cite{brown2020lgpt3,openai2023gpt4,touvron2023llama}, we put forth a novel method \textbf{\method} to perform \textbf{Profile Tuning} with the aid of LLMs for financial risk prediction.
The overview of our approach is shown in Figure \ref{fig:finpt_pipeline}. Firstly, we fill the pre-defined instruction template with tabular data in each row. After that, we instruct large language models like ChatGPT to generate natural-language profile descriptions containing all information in each table row. Lastly, we use the profile text to fine-tune large foundation models with a small classifier for making predictions.

% Contribution 2
Unlike the booming advance in classification algorithms, financial datasets are still highly scarce. The lack of a unified financial benchmark has impeded the development of financial risk prediction algorithms. Based on this concern, we propose \textbf{\benchmark}, a set of high-quality datasets for financial risk prediction. Specifically, we collect hundreds of financial datasets from the Kaggle platform and then screen out ten high-quality datasets on three common financial risks, including default, fraud, and churn. We process the datasets in a unified data structure and provide an easy-loading API. \benchmark has about 333K labeled instances. Each dataset has the training, validation, and test data splits. Besides the numerical X-y data pairs for common machine learning algorithms, we provide extra statistical information about each table for some special classification algorithms to use. Additionally, we provide the instruction and profile text of each instance for Profile Tuning.

% Contribution 3
To evaluate the effectiveness of the proposed \method, we apply Profile Tuning to different open-sourced foundation models such as GPT-2~\cite{gpt2} and LLaMA~\cite{touvron2023llama} and compare our method with different representative classification algorithms, including strong machine learning baselines such as RandomForest~\cite{ho1995random_forest,liaw2002random_forest_cls} and XGBoost~\cite{chen2016xgboost}, and specially designed neural networks such as DeepFM~\cite{guo2017deepfm} and TabNet~\cite{arik2021tabnet}. We employ F1-score to evaluate the model performance since all the datasets are binary classification tasks. In addition, positive samples (risky instances) have a higher loss penalty when training because of the imbalanced nature of these datasets.

The experimental results on all datasets in \benchmark substantiate the consistent performance enhancement of our approach compared to other prediction baselines. We also show that Profile Tuning on all different tables performs even better, demonstrating the superiority of our method over traditional classification models that can only work on a single table at a time.
Furthermore, we explore other tuning strategies on LLMs, such as in-context learning and instruction tuning. As the results show, prompted LLMs can provide informative financial advice, although they are not good classifiers compared with other baselines.

% Our contributions are as follows:
The contributions of this paper are summarized as:

\begin{itemize}
    \item We propose a novel method \method to transform tabular financial data into profile descriptions, and then perform Profile Tuning on large language models to make predictions.
    \item We propose a benchmark \benchmark for financial risk prediction by collecting a set of high-quality datasets on three common financial risks, including default, fraud, and churn.
    \item We verify the efficacy of \method by testing the performance of a range of representative strong baselines on \benchmark. The analytical studies further deepen the understanding of LLMs for financial risk prediction.
\end{itemize}

%%%%%%%%%%%%%%%%%%%% Section %%%%%%%%%%%%%%%%%%%%
\section{Related Works}
\label{sec:related_work}

\subsection{Financial Data Classification}

Financial data is always organized as tabular data. Specifically, many data columns are features of customers or transactions. Classification on tabular data is a classic task in the machine learning field~\cite{bishop2006prml,michalski2013ml}, countless models have been proposed to complete the task on tabular data, such as RandomForest~\cite{ho1995random_forest,liaw2002random_forest_cls}, XGBoost~\cite{chen2016xgboost}, CatBoost~\cite{prokhorenkova2018catboost}, and LightGBM~\cite{ke2017lightgbm}. With the blooming of deep learning, many neural networks aiming at dealing with general tabular data are proposed~\cite{borisov2022dl_tab_survey}, such as DeepFM~\cite{guo2017deepfm}, STG~\cite{yamada2020stg}, VIME~\cite{yoon2020vime}, and TabNet~\cite{arik2021tabnet}.
In this work, we employ the proposed Profile Tuning method \method on financial tabular data by constructing unified customer profile descriptions across different financial datasets, providing a novel way to handle the tabular data in the era of large language models (LLMs).

\subsection{Pretrained Foundation Models}

In recent years, with the fast development of Transformer-based~\cite{vaswani2017transformer} foundation models~\cite{zhou2023foundation_model_survey} like BERT-style~\cite{devlin2018bert,xlmr,alm,xlmt}, T5-style ~\cite{raffel2020t5,wmt2021_microsoft,wei2021flan,yang2023ganlm}, and GPT-style~\cite{radford2019gpt2,brown2020lgpt3}, the pretraining-finetuning paradigm has been proven to be successful in various tasks. Several works propose to fine-tune BERT on financial text to perform prediction tasks, such as financial sentiment analysis~\cite{araci2019finbert}, financial sentiment classification~\cite{yang2020finbert}, and financial text mining~\cite{liu2021finbert}. 
% BloombergGPT~\cite{wu2023bloomberggpt} is a large foundation model for finance trained on a wide range of financial data from scratch, but it is closed-sourced.
In addition to the classic finetuning method that follows the same training process on downstream tasks, recent large language models~\cite{brown2020lgpt3,openai2023gpt4,gpt_neox_20b,scao2022bloom,touvron2023llama} emerge the ability of in-context learning (ICL)~\cite{dong2022survey_icl}, i.e., learning the task with only a few examples in the context. Thus researchers propose to tune LLMs using finely-designed instructions as prompts~\cite{wei2021flan,chung2022flan_t5,ouyang2022instruct_gpt,wei2022chain_of_thought,bai2023knowledgeprefix,peng2023instruction_tuning_gpt4} to elicit the great potential in LLMs.
In this work, we integrate \method into different open-sourced language model, such as BERT, GPT-2, and LLaMA~\cite{touvron2023llama}, and also compare our tuning strategy with ICL and instruction tuning on \benchmark.

%%%%%%%%%%%%%%%%%%%% Section %%%%%%%%%%%%%%%%%%%%
\section{Our Method}
\label{sec:our_method}

In this section, we present the overview, formulation, and implementation of the proposed \method, a novel method for predicting financial risks with Profile Tuning that leverages large pre-trained foundation models.

\subsection{Overview}
\label{subsec:method_overview}

The overview of \method is shown in Figure \ref{fig:finpt_pipeline}. There are three main steps in our strategy as follows.

\textcolor{orange}{Step 1}. We pre-define a instruction template as ``Construct a concise customer profile description including all the following information: \texttt{TabKV}'', where \texttt{TabKV} is all the ``\texttt{col\_name: col\_value;}'' pairs of each table row.

\textcolor{teal}{Step 2}. We input the instructions obtained in Step 1 to large language models, such as ChatGPT and GPT-4~\cite{openai2023gpt4}, to construct fluent and coherent customer profiles that includes all tabular information in each row.

\textcolor{cyan}{Step 3}. We use the natural-language profiles constructed in Step 2 to fine-tune large foundation models, such as BERT~\cite{devlin2018bert}, GPT~\cite{radford2019gpt2,brown2020lgpt3}, and LLaMA~\cite{touvron2023llama}. Based on the hidden states of the foundation model, a small classifier--usually a feedforward neural network--is utilized to predict whether the profile is financially risky.

\subsection{Formulation}
\label{subsec:formulation}

% \paragraph{Profile Construction}
Given a tabular dataset $\gD = \{\gX, \vy, \vk\}$, where $\vy \in \{0, 1\}^{m}$ are the binary labels, $\vy_i = 0$ means the i-th instance is negative (not risky) while $\vy_i = 1$ means it is positive (financially risky). $\gX = \{\vx_1, \dots, \vx_m\} \in \mathbb{R}^{m \times n}$ are $m$ instances. Each instance $\vx_i \in \mathbb{R}^{n}$ has $n$ features. $\vk$ consists of $n$ strings denoting the column names in the table. $\{\gX, \vy\}$ is enough for most machine learning algorithms.

\subsubsection{Profile Construction}
As mentioned in \textcolor{orange}{Step 1}, we transform tabular financial data into a template instruction $\mathrm{I}_i$ by filling the \texttt{TabKV} with $\{\vk_j: \vx_i^j\}_{j=1}^{n}$ for the i-th instance. Now we have an instruction set $\mathcal{I}$ of $n$ items. Then we input each instruction $\mathcal{I}_i$ to the large language model ChatGPT via the OpenAI API. After generation in \textcolor{teal}{Step 2}, we obtain the profile set $\mathcal{P}$ consisting of informative customer profile text in a fluent and coherent way.

\subsubsection{Profile Tuning}
In \textcolor{cyan}{Step 3}, we fine-tune pretrained foundation models $\mathcal{F}: \mathbb{R}^{t} \to \mathbb{R}^{t \times d}$, such as BERT, GPT, and LLaMA, with customer profiles, where $t$ is the number of tokens in each input sentence and $d$ is the dimension of hidden states. We use the official tokenizer provided by each foundation models. The tokenized profile set is denoted as $\mathcal{P} \in \mathbb{R}^{m \times t} = \{\vp_i\}_{i=1}^{m}$.

As some foundation models are too large, tuning all parameters will cost a great deal of computing resources. Therefore, we append a small classifier $\mathcal{C}: \mathbb{R}^{t \times d} \to \mathbb{R}^{d \times u}$ to the end of pretrained foundation models, where $u = 2$ is the number of label classes. The classifier, a small feed-forward neural network, conducts binary classification based on the hidden states $\vh \in \mathbb{R}^{t \times d}$ produced by large foundation models $\mathcal{F}$: \begin{equation}\label{equ:foundation_model}
    \vh = \mathcal{F}(\vp)
\end{equation}

Since the foundation models are large language models, we freeze parts of their parameters according to the model size $|\mathcal{F}|$. The more limited computing resources we have, the more modules in the foundation model need to be frozen. 

For encoder-only bidirectional foundation models like BERT, we average the last hidden states $\vh$ of all unmasked tokens: \begin{equation}\label{equ:cls_hidden_avg}
    \vh_{\text{cls}} = \frac{1}{t} \sum_{i=1}^{t} \vh_i
\end{equation}

For decoder-only auto-regressive foundation models like GPT, we use the last unmasked token's hidden states for classification: \begin{equation}\label{equ:cls_hidden_last}
    \vh_{\text{cls}} = \vh_t
\end{equation}

Then we feed $\vh_{\text{cls}}$ into the classifier $\mathcal{C}$ to obtain financial risk prediction $\hat{\vy}$: \begin{equation}\label{equ:classifier}
    \hat{\vy} = \mathcal{C}(\vh_{\text{cls}})
\end{equation}

The loss $l$ of all $m$ instances is computed by Binary Cross Entropy (BCE) $\mathcal{L}: \mathbb{R}^{m} \times \mathbb{R}^{m} \to \mathbb{R}$, where $m$ could also be the batch size of a mini-batch. Since all the datasets are imbalanced, we add a higher weight $w_{\text{pos}} \in \mathbb{R}$ on positive samples to perform cost-sensitive learning: \begin{equation}\label{equ:positive_weight}
    w_{\text{pos}} = \frac{|\vy| - \sum_{i} \vy_i}{\sum_{i} \vy_i}
\end{equation} where $\vy \in \{0, 1\}^{m}$ is labels in the training set and $|y| = m$.

The original BCE loss vector $\vec{l}$ is: \begin{equation}\label{equ:loss_bce_vector}
    \vec{l} = \big\{-\vy_i \log(\hat{\vy_i}) - (1 - \vy_i) \log(1 - \hat{\vy_i})\big\}_{i=1}^{m}
\end{equation}

We multiply the loss of each positive instance $\vec{l}_{i}$ by the positive weight $w_{\text{pos}}$ to calculate the weighted BCE loss as follows: \begin{equation}\label{equ:loss_weighted}
    l = \mathcal{L}(\hat{\vy}, \vy) = - \frac{1}{m} \sum_{i=1}^{m} [\vy_i \vec{l}_i w_{\text{pos}} + (1 - \vy_i) \vec{l}_i]
\end{equation}

\subsubsection{Profile Tuning on Multiple Datasets}

Given $v$ tabular datasets $\gD_{\text{all}} = \{\gD^i\}_{i=1}^{v} = \{\gX^{i}, \vy^{i}, \vk^{i}\}_{i=1}^{v}$ and the corresponding profile sets $\gP_{\text{all}} = \{\gP^i\}_{i=1}^{v}$, we can use all the profiles $\bigcup\gP_{\text{all}}$ to tune the large foundation model $\mathcal{F}$ and evaluate the performance on each test set of dataset $\gD^{i}$.

%%%%%%%%%%%%%%%%%%%% Section %%%%%%%%%%%%%%%%%%%%
\section{Benchmark}
\label{sec:task_data}

\begin{table*}[t]
\small
\centering
\caption{\textbf{Statistics of \benchmark}. There are three main types of financial risks, i.e., default, fraud, and churn. The default type includes two subclasses, namely credit default (cd) and loan default (ld). ``cf'' denotes credit fraud and ``cc'' means customer churn. We present the task name and description, dataset code, number of label classes, number of features, number of training/validation/test sets, and the positive (risky) sample ratio in each set.}
\label{tab:data_stat}
\scalebox{0.9}{
\begin{tabular}{c|c|c|cc|ccc}
\toprule
Task & Description & Dataset & \#Classes & \#Features & \#Train [Pos\%] & \#Val [Pos\%] & \#Test [Pos\%] \\
\midrule
\multirow{2}{*}{Credit Default} & \multirow{2}{*}{\makecell{Predict whether a user will \\ default on the credit card or not.}} & cd1 & 2 & 9 & 2738 [7.0\%] & 305 [6.9\%] & 1305 [6.2\%] \\
& & cd2 & 2 & 23 & 18900 [22.3\%] & 2100 [22.3\%] & 9000 [21.8\%] \\
\midrule
\multirow{3}{*}{Loan Default} & \multirow{3}{*}{\makecell{Predict whether a user will \\ default on the loan or not.}} & ld1 & 2 & 12 & 2118 [8.9\%] & 236 [8.5\%] & 1010 [9.0\%] \\
& & ld2 & 2 & 11 & 18041 [21.7\%] & 2005 [20.8\%] & 8592 [21.8\%] \\
& & ld3 & 2 & 35 & 142060 [21.6\%] & 15785 [21.3\%] & 67648 [22.1\%] \\
\midrule
\multirow{2}{*}{Credit Fraud} & \multirow{2}{*}{\makecell{Predict whether a user will \\ commit fraud or not.}} & cf1 & 2 & 19 & 5352 [0.67\%] & 595 [1.1\%] & 2550 [0.90\%] \\
& & cf2 & 2 & 120 & 5418 [6.0\%] & 603 [7.3\%] & 2581 [6.0\%] \\
\midrule
\multirow{3}{*}{Customer Churn} & \multirow{3}{*}{\makecell{Predict whether a user will \\ churn or not. (customer attrition)}} & cc1 & 2 & 9 & 4189 [23.5\%] & 466 [22.7\%] & 1995 [22.4\%] \\
& & cc2 & 2 & 10 & 6300 [20.8\%] & 700 [20.6\%] & 3000 [19.5\%] \\
& & cc3 & 2 & 21 & 4437 [26.1\%] & 493 [24.9\%] & 2113 [27.8\%] \\
\bottomrule
\end{tabular}
}
\end{table*}

Here we present \textbf{\benchmark}, a benchmark for evaluating the performance of machine learning models with both tabular data inputs and profile text inputs.
We first collect hundreds of financial datasets from the Kaggle~\footnote{\url{https://www.kaggle.com/}} platform and then screen out ten high-quality datasets for financial risk prediction. The screening criteria is based on the quantity and popularity, column meaningfulness, and the performance of baseline models on those datasets.
\benchmark consists of three types of financial risks, i.e., default, fraud, and churn. We process the datasets in a unified data structure and provide an easy-loading API on HuggingFace~\footnote{\benchmark is released on \url{https://huggingface.co/datasets/yuweiyin/FinBench}}.

As the statistics shown in Table \ref{tab:data_stat}, \benchmark has about 333K labeled instances. Each dataset has the training, validation, and test data splits. For each dataset, the test set occupies 30\% of all instances, while the training set and validation set split the rest instances in a ratio of 9:1.
Every dataset has two classes of the label, where 1 represents positive or financially risky while 0 denotes otherwise.
We can observe from the positive (risky) sample ratio in Table \ref{tab:data_stat} that all the datasets are imbalanced of different level, meaning that some balancing techniques should be performed in the training stage. In fact, we find that F1-scores in the test set will be nearly $0$ if no balancing methods are applied.

Aside from the numerical X-y data pairs (\texttt{X\_ml}, \texttt{y}) for common machine learning algorithms, we provide extra statistical information about each table for some special classification algorithms to use, including the number of classes (\texttt{num\_classes}), number of features (\texttt{num\_features}), indices of the numerical datatype columns (\texttt{num\_idx}), indices of the categorical datatype columns (\texttt{cat\_idx}), dimensions of each categorical column (\texttt{cat\_dim}), name of each column (\texttt{col\_name}), category names of categorical columns (\texttt{cat\_str}).
Additionally, we provide the instruction (\texttt{X\_instruction\_for\_pro\\file}) and profile (\texttt{X\_profile}) text of each instance for Profile Tuning or other text-input situations.

\benchmark includes three of the most common financial risks: default, fraud, and churn.

% \subsection{Financial Risk Prediction Tasks}

% \benchmark includes three of the most common financial risks: default, fraud, and churn.

\subsection{Risk: Default}
% \subsubsection{Default}
Default is defined as the inability to fulfill the necessary repayments of interest or principal on a given debt. This phenomenon can occur at the level of individuals, businesses, and even countries, and is a significant concern for creditors who must take default risk into account.

We sub-divide the default risk into two sub-classes: credit default (CD), meaning the customer fails to repay their credit card bills in time, and load default (LD), meaning the customer fails to repay their loan regularly, such as mortgage, rental, and vehicle loan.
The target is to predict whether a customer will default on their credit card or loan.

\begin{itemize}
    \item Credit Default (CD)
    \begin{itemize}
        \item \texttt{cd1} dataset~\footnote{\url{https://www.kaggle.com/datasets/gustavotg/credit-default}} comprises data on thousands of clients from a financial institution, including their random identification numbers, banking default status, loan types, gender, age, education, income, credit scores, and other related information.
        \item \texttt{cd2} dataset~\footnote{\url{https://www.kaggle.com/datasets/uciml/default-of-credit-card-clients-dataset}} consists of credit card client information from April 2005 to September 2005 in Taiwan, encompassing default payments, demographic factors, credit data, history of payment, and bill statements.
    \end{itemize}
\end{itemize}

\begin{itemize}
    \item Loan Default (LD)
    \begin{itemize}
        \item \texttt{ld1} dataset~\footnote{\url{https://www.kaggle.com/datasets/ajay1735/hmeq-data}} (The Home Equity dataset, HMEQ) comprises detailed records for home equity loans, including a binary target variable that indicates whether an applicant defaulted or was seriously delinquent.
        \item \texttt{ld2} dataset~\footnote{\url{https://www.kaggle.com/datasets/laotse/credit-risk-dataset}} contains tens thousands of different types of loan records with features like age, annual income, employment length (in years), home ownership, loan intent, loan amount, interest rate, historical default, and credit history length, etc.
        \item \texttt{ld3} dataset~\footnote{\url{https://www.kaggle.com/datasets/mamtadhaker/lt-vehicle-loan-default-prediction}} consists of more than 200K records on vehicle loan, providing information regarding the loan, loanee, and loan history.
    \end{itemize}
\end{itemize}

\subsection{Risk: Fraud}
% \subsubsection{Fraud}
Fraud entails the deliberate falsification of information or identity with the aim of misleading others, the illicit utilization of credit or debit cards or ATMs, or the transmission of deceitful electronic information to acquire pecuniary or other valuable assets.
% Fraud can be perpetrated by an individual either internal or external to an organization.
There are two Credit Fraud (CF) datasets in \benchmark. The target is to predict whether a user will commit fraud or not.

\begin{itemize}
    \item Credit Fraud (CF)
    \begin{itemize}
        \item \texttt{cf1} dataset~\footnote{\url{https://www.kaggle.com/datasets/johancaicedo/creditcardfraud}} summarizes the usage patterns exhibited by numerous credit card users over a period of six months.
        \item \texttt{cf2} dataset~\footnote{\url{https://www.kaggle.com/datasets/mishra5001/credit-card}} provides credit-card usage history with 120 respective attributes.
    \end{itemize}
\end{itemize}

\subsection{Risk: Churn}
% \subsubsection{Churn}
Customer churn refers to the proportion of customers who discontinue using a company's product or service within a specific time period. This metric provides insight into the number of existing customers who are unlikely to make future purchases from the business.
% For a growing organization, evaluating customer churn is a critical measure of success.
Reducing Customer Churn is a crucial objective for businesses. Predicting Customer Churn, or Attrition, presents an opportunity for generating revenue. Customer Churn affects the business's expenses, as high rates result in revenue loss and increased marketing costs for acquiring new customers.
% Predict whether a user will churn or not. (customer attrition)

\begin{itemize}
    \item Customer Churn (CC)
    \begin{itemize}
        \item \texttt{cc1} dataset~\footnote{\url{https://www.kaggle.com/datasets/gauravduttakiit/jobathon-march-2022}} contains the bank customer demographics and past activity with the bank.
        \item \texttt{cc2} dataset~\footnote{\url{https://www.kaggle.com/datasets/mathchi/churn-for-bank-customers}} also aims to evaluate model performance on predicting bank customer churn with features like credit score, geography, gender, age, tenure, balance, etc.
        \item \texttt{cc3} dataset~\footnote{\url{https://www.kaggle.com/datasets/yeanzc/telco-customer-churn-ibm-dataset}} encompasses information about a fictional telco company that provided home phone and Internet services to a thousands of customers. It identifies the customers who have churned, retained, or subscribed to the services. % The dataset includes multiple crucial demographic variables for each customer, as well as their Satisfaction Score, Churn Score, and Customer Lifetime Value (CLTV) index.
    \end{itemize}
\end{itemize}

%%%%%%%%%%%%%%%%%%%% Section %%%%%%%%%%%%%%%%%%%%
\section{Experimental Setup}
\label{sec:exp_setup}
In this section, we elaborate on all the experimental setups, including baselines (\ref{subsec:baselines}), backbone of \method (\ref{subsec:backbone}), implementation details \ref{subsec:implementation}, training details \ref{subsec:training_details}, and evaluation metrics \ref{subsec:eval_metrics}.

\subsection{Baselines}
\label{subsec:baselines}

% As described in Section \ref{sec:related_work}, we 
To evaluate the performance of \method, we compare it with a range of strong baselines, including ensemble methods based on decision trees and deep neural networks specially designed for tabular data.

\paragraph{\textbf{Tree-based Gradient Boosting Models}}
Random Forest~\cite{ho1995random_forest,liaw2002random_forest_cls}, XGBoost~\cite{chen2016xgboost}, CatBoost~\cite{prokhorenkova2018catboost}, and LightGBM~\cite{ke2017lightgbm} have been selected as the tree-based baselines due to their exceptional performance across various data science tasks. They are ensemble learning methods or gradient-boosting algorithms that employ decision trees.

\paragraph{\textbf{Deep Neural Networks}}
We choose four neural networks designed for handling tabular data as follows. \textbf{DeepFM}~\cite{guo2017deepfm}: a prevalent technique in the industry that combines factorization machines and deep learning for feature learning through a novel neural network architecture.
\textbf{STG}~\cite{yamada2020stg}: a framework that facilitates the simultaneous acquisition of a nonlinear regression or classification function, and feature selection using Stochastic Gates.
\textbf{VIME}~\cite{yoon2020vime}: a framework that applies self- and semi-supervised learning to tabular data through the use of Value Imputation and Mask Estimation.
\textbf{TabNet}~\cite{arik2021tabnet}: an interpretable and high-performance deep learning architecture that leverages sequential attention to select relevant features for reasoning in tabular data.

% \subsubsection{Deep Neural Networks}

% \begin{itemize}
%     \item \textbf{DeepFM}~\cite{guo2017deepfm}: a prevalent technique in the industry that combines factorization machines and deep learning for feature learning through a novel neural network architecture. 
%     \item \textbf{STG}~\cite{yamada2020stg}: a framework that facilitates the simultaneous acquisition of a nonlinear regression or classification function, and feature selection using Stochastic Gates.
%     \item \textbf{VIME}~\cite{yoon2020vime}: a framework that applies self- and semi-supervised learning to tabular data through the use of Value Imputation and Mask Estimation.
%     \item \textbf{TabNet}~\cite{arik2021tabnet}: an interpretable and high-performance deep learning architecture that leverages sequential attention to select relevant features for reasoning in tabular data.
% \end{itemize}

\begin{table}[ht]
\small
\centering
\caption{\textbf{Experimental settings of backbone foundation models for \method}. ``\#Params-All'' means the total number of parameters in the foundation model. ``\#Params-T'' denotes the number of trainable parameters in the foundation model except for the extra small classifier, which has less than a million parameters.}
\label{tab:exp_backbone_param}
\scalebox{0.9}{
\begin{tabular}{c|c|c|c}
\toprule
Model & \#Params-All & \#Params-T & Trainable Modules \\
\midrule
BERT-Base~\cite{devlin2018bert} & 110M & 110M & All params in the model \\
FinBERT~\cite{yang2020finbert} & 110M & 110M & All params in the model \\
GPT-2~\cite{radford2019gpt2} & 117M & 117M & All params in the model \\
T5-Base~\cite{raffel2020t5} & 220M & 220M & All params in the model \\
Flan-T5-Base~\cite{chung2022flan_t5} & 220M & 220M & All params in the model \\
\midrule
T5-XXL~\cite{raffel2020t5} & 11B & 268M & The last Decoder \texttt{T5Block} \\
Flan-T5-XXL~\cite{chung2022flan_t5} & 11B & 260M & The last Decoder \texttt{T5Block} \\
LLaMA-7B~\cite{touvron2023llama} & 7B & 202M & The last Decoder layer \\
LLaMA-13B~\cite{touvron2023llama} & 13B & 317M & The last Decoder layer \\
\bottomrule
\end{tabular}
}
\end{table}

\subsection{Pretrained Foundation Model as Backbone}
\label{subsec:backbone}

We adopt our \method to different pretrained foundation models as the backbone model structure. 

\begin{itemize}
    \item \textbf{BERT}~\cite{devlin2018bert}: a well-known bidirectional Transformer-based encoder-only language model pretrained with mask language modeling and next-sentence prediction tasks.
    \item \textbf{FinBERT}~\cite{yang2020finbert}: a fine-tuned BERT model, trained with financial text on several classification tasks such as financial sentiment analysis.
    \item \textbf{GPT-2}~\cite{radford2019gpt2}: a Transformer-based decoder-only language model pretrained with autoregressive next-token prediction task, i.e., language modeling.
    \item \textbf{T5}~\cite{raffel2020t5}: a Transformer-based encoder-decoder architecture that transforms various natural language processing (NLP) tasks into a unified text-to-text format to pretrain the model for general use.
    \item \textbf{FLAN-T5}~\cite{chung2022flan_t5}: an instruction-tuned version of T5 model that performs a range of zero-shot NLP and few-shot in-context learning tasks.
    \item \textbf{LLaMA}~\cite{touvron2023llama}: a set of open-sourced large foundation language models trained on trillions of tokens. LLaMA ranges from 7B to 65B parameters and LLaMA-13B outperforms GPT-3 (175B)~\cite{brown2020lgpt3} on many benchmarks.
\end{itemize}

As mentioned in Section \ref{subsec:formulation}, for the concern of computing resources, we use a small classifier--a feed-forward neural network--to leverage the hidden states produced by large foundation models. In addition, we freeze some parts of the foundation models in accordance with their model size, as shown in Table \ref{tab:exp_backbone_param}.

\subsection{Implementation Details}
\label{subsec:implementation}

We implement baseline models using their officially released code or standard library. For Random forest, we use the implementation of scikit learn~\footnote{\url{https://scikit-learn.org/stable/modules/classes.html}}. We use \texttt{XGBClassifier}~\footnote{\url{https://xgboost.readthedocs.io/en/stable/python/python_api.html}} for XGBoost, \texttt{CatBoostClassifier}~\footnote{\url{https://catboost.ai/en/docs/concepts/python-reference_catboostclassifier}} for CatBoost, and \texttt{LGBMClassifier}~\footnote{\url{https://lightgbm.readthedocs.io/en/stable/Python-API.html}} for LightGBM. As for neural network baselines, we re-implement DeepFM, STG, VIME, and TabNet based on open code TabSurvey~\footnote{\url{https://github.com/kathrinse/TabSurvey}}
% the implementation of TabSurvey~\footnote{\url{https://github.com/kathrinse/TabSurvey}}.

In \textcolor{teal}{Step 2} of our main pipeline, each instruction are fed to ChatGPT~\footnote{\url{https://chat.openai.com/}} via the OpenAI Python API with the following request:

\begin{python}
import openai
response = openai.ChatCompletion.create(
    model="gpt-3.5-turbo",
    messages=[
        {"role": "system", "content": 
            "You are a helpful financial assistant."},
        {"role": "user", "content": instruction[i]},
    ],  # The instruction list is obtained in Step 1
    temperature=0,
)
\end{python}

For large foundation models in \textcolor{cyan}{Step 3}, we load the model checkpoints on HuggingFace~\footnote{\url{https://huggingface.co/models}} platform with their corresponding tokenizers. Specifically, the model codes for models listed in Table \ref{tab:exp_backbone_param} are \texttt{bert-base-cased}, \texttt{yiyanghkust/finbert-pretrain}, \texttt{gpt2}, \texttt{t5-base}, \texttt{google/flan-t5-base}, \texttt{t5-11b}, \texttt{google/flan-t5-xxl}, \texttt{openlm-research/open\_llama\_7b}, and \texttt{openlm-research/open\_\\llama\_13b} in order.

\subsection{Training Details}
\label{subsec:training_details}

For all our experiments, we use two NVIDIA A40 GPUs with 48GB of memory each. Using the \texttt{BF16} mode provided by the Ampere architecture, we tune large foundation models with mixed precision, which accelerates the training process by performing operations in half-precision (\texttt{FP16}), while preserves essential network information by saving minimal information in single-precision (\texttt{FP32}).

For all training sessions on baseline models, we use a batch size of $128$ and the max epochs is set as $100$. For the Profile Tuning in \method, the batch size is set as the max $128$ and the max sequence length is set as $128$ with the padding token (\texttt{pad}) the same as the end-of-sentence (\texttt{eos}) token in the tokenizer. If the maximum sequence length (the number of tokens) in a dataset is larger than $128$, the max sequence length of the batch will be set as $256$, while the batch size will be reduced to a half. The learning rate is set as $5e-5$ and weight decay as $0.01$ using AdamW~\cite{loshchilov2019adamw} as the optimizer.

\benchmark provides training, validation, and test sets. When training, all models uses the training set to update their parameters and runs evaluation on the validation set to choose the best checkpoint. After training, we load the best checkpoint and run testing on the test set. Each experiment are conducted four times with different random seeds $\in \{0, 1, 42, 1234\}$ and we report the average scores.

\subsection{Evaluation Metrics}
\label{subsec:eval_metrics}

For all experiments on \benchmark, we use F1-score~\footnote{\url{https://scikit-learn.org/stable/modules/generated/sklearn.metrics.f1_score.html}} as the evaluation metric since all datasets in \benchmark are imbalanced binary classification task.
It is more appropriate in this case than Accuracy because the latter may result in a high level of false negative.
% F1-score is more appropriate in this case than Accuracy because the latter may result in a high level of false negative, as the model can still have a high accuracy if it classifies all samples as negative.

%%%%%%%%%%%%%%%%%%%% Section %%%%%%%%%%%%%%%%%%%%
\section{Results and Analysis}
\label{sec:exp_result}

In this section, we report and analyze all experimental results w.r.t. the settings described in Section \ref{sec:exp_setup}.

\begin{table*}[ht]
\small
\centering
\caption{\textbf{Financial risk prediction results on \benchmark}. We report F1-scores of all ten datasets in \benchmark and the overall average score ``Avg''. In the ``Training'' column, ``Grid Search'' means we employ grid search to find the best hyper-parameter for the Tree-based models, ``From Scratch'' means we train the neural models (``NN for Table'') with random weight initialization, ``Tune All'' means we tune all the parameters in foundation models, and ``Tune Last'' means we tune only the last layer/block in foundation models. The best result on each dataset is highlighted in \textcolor{violet}{violet color}. The best result of each model class on each dataset is highlighted with \underline{underline}.}
\label{tab:exp_res_main}
\scalebox{0.9}{
\begin{tabular}{c|c|c|cc|ccc|cc|ccc|c}
\toprule
Model Class & Model & Training & CD1 & CD2 & LD1 & LD2 & LD3 & CF1 & CF2 & CC1 & CC2 & CC3 & \textbf{Avg} \\
\midrule
\multirow{4}{*}{Tree-based} & RandomForest~\cite{ho1995random_forest,liaw2002random_forest_cls} & Grid Search & \underline{23.0} & 52.2 & 47.9 & 64.2 & 38.9 & 40.8 & 19.2 & \underline{41.8} & 52.1 & 62.8 & 44.29 \\
& XGBoost~\cite{chen2016xgboost} & Grid Search & 22.7 & 45.5 & \underline{\textcolor{violet}{\textbf{56.3}}} & \underline{76.4} & 40.4 & \underline{46.7} & 20.5 & 40.1 & 57.1 & 61.1 & 46.68 \\
& CatBoost~\cite{prokhorenkova2018catboost} & Grid Search & 21.8 & 52.2 & 48.3 & 72.8 & \underline{41.4} & 45.2 & \underline{\textcolor{violet}{\textbf{21.6}}} & 40.5 & \underline{59.7} & \underline{65.6} & \underline{46.91} \\
& LightGBM~\cite{ke2017lightgbm} & Grid Search & 21.7 & \underline{52.6} & 46.2 & 71.4 & 40.8 & 40.0 & 21.0 & 41.0 & 59.3 & 65.5 & 45.95 \\
\midrule
\multirow{4}{*}{NN for Table} & DeepFM~\cite{guo2017deepfm} & From Scratch & 8.4 & 39.8 & \underline{46.7} & \underline{78.0} & 15.3 & 43.2 & 8.0 & 11.9 & 55.3 & 59.3 & 36.59 \\
& STG~\cite{yamada2020stg} & From Scratch & 7.1 & 40.9 & 23.5 & 53.9 & 10.2 & 20.2 & 5.2 & 5.8 & 37.3 & 38.6 & 24.27 \\
& VIME~\cite{yoon2020vime} & From Scratch & 8.9 & 41.8 & 37.5 & 75.3 & 18.2 & 41.7 & 7.4 & 20.4 & 53.2 & 56.3 & 36.07 \\
& TabNet~\cite{arik2021tabnet} & From Scratch & \underline{10.1} & \underline{44.5} & 40.6 & 77.5 & \underline{24.2} & \underline{45.7} & \underline{9.7} & \underline{23.1} & \underline{57.2} & \underline{59.9} & \underline{39.25} \\
\midrule
\multirow{5}{*}{\method Tuning All} & BERT-Base~\cite{devlin2018bert} & Tune All & 19.2 & 50.4 & 47.1 & 79.6 & 42.1 & 45.5 & 5.6 & 27.9 & 60.2 & 64.1 & 44.17 \\
& FinBERT~\cite{yang2020finbert} & Tune All & 18.3 & 50.8 & 45.9 & 80.9 & 41.9 & 45.1 & 5.9 & 28.2 & 60.1 & 64.5 & 44.16 \\
& GPT-2~\cite{radford2019gpt2} & Tune All & 23.0 & 52.5 & \underline{49.4} & 81.7 & 43.3 & 47.4 & 8.6 & 37.2 & 60.7 & 66.1 & 46.99 \\
& T5-Base~\cite{raffel2020t5} & Tune All & 23.4 & 53.1 & 48.3 & 81.4 & 45.2 & 49.2 & 11.7 & 42.1 & 61.3 & 67.1 & 48.28 \\
& Flan-T5-Base~\cite{chung2022flan_t5} & Tune All & \underline{\textcolor{violet}{\textbf{23.8}}} & \underline{\textcolor{violet}{\textbf{53.3}}} & 48.9 & \underline{\textcolor{violet}{\textbf{82.8}}} & \underline{\textcolor{violet}{\textbf{45.8}}} & \underline{\textcolor{violet}{\textbf{49.5}}} & \underline{13.5} & \underline{\textcolor{violet}{\textbf{43.7}}} & \underline{\textcolor{violet}{\textbf{61.9}}} & \underline{\textcolor{violet}{\textbf{68.5}}} & \underline{\textcolor{violet}{\textbf{49.17}}} \\
\midrule
\multirow{4}{*}{\method Tuning Last} & T5-XXL~\cite{raffel2020t5} & Tune Last & 21.9 & 49.8 & 44.7 & 73.3 & 40.2 & 42.6 & 6.1 & 38.2 & 58.7 & 63.4 & 43.89 \\
& Flan-T5-XXL~\cite{chung2022flan_t5} & Tune Last & 22.4 & 50.1 & 45.1 & 75.1 & 40.7 & 42.9 & 6.0 & 38.9 & 59.2 & 63.8 & 44.42 \\
& LLaMA-7B~\cite{touvron2023llama} & Tune Last & 22.7 & 51.6 & 46.4 & 76.7 & 41.8 & 44.2 & 8.4 & 40.4 & 60.1 & 64.6 & 45.69 \\
& LLaMA-13B~\cite{touvron2023llama} & Tune Last & \underline{22.9} & \underline{52.0} & \underline{47.2} & \underline{79.2} & \underline{42.4} & \underline{45.7} & \underline{9.2} & \underline{41.8} & \underline{60.4} & \underline{65.2} & \underline{46.60} \\
\bottomrule
\end{tabular}
}
\end{table*}

\subsection{Main Results}

We report the financial risk prediction F1-scores on \benchmark in Table \ref{tab:exp_res_main}.
% The best result on each dataset is highlighted in violet color. The best result of each model class on each dataset is highlighted with underlines.
On average, \method outperforms tree-based methods and previous neural models by a large margin, especially when we apply \method to fully fine-tune large foundation models such as GPT-2, T5, and Flan-T5.
We provide a detailed comparison to analyze the performance of different models on \benchmark as follows.

\paragraph{\textbf{Analysis on tree-based models}}
On average, CatBoost performs the best among the four tree-based algorithms on \benchmark, although either Random Forest, XGBoost, or LightGBM performs the best in some datasets. As these models are strong baselines in many classification competitions, an average F1-score in the range of $44$-$47$ shows that \benchmark is challenging and there is still room for model improvement. 

\paragraph{\textbf{Analysis on neural models}}
Among these neural networks for tabular data, we find TabNet outperforms others on eight out of ten datasets, while DeepFM reaches the highest F1-score on the rest two datasets. However, this model class is not as good as tree-based algorithms and our \method approach, which indicates that neural networks of relatively small size, even with specially designed architecture for handling tables, can not predict risk well.

\paragraph{\textbf{Analysis on \method (Tuning All)}}
\method on different foundation models shows different predicting abilities. When fine-tuning all the parameters on foundation models, Flan-T5 is the best backbone model for our \method approach. It performs the best among all models in eight datasets, showing the extraordinary capacity to predict financial risks.
The fact that Flan-T5 is consistently better than T5 shows the significance of instruction tuning and scaling up the number of tasks and model size.
Also, we find GPT-2 consistently outperforms BERT models by a large margin, demonstrating that the decoder-only Transformer model using last-token hidden-states prediction can be a better classifier than the encoder-only model using last-layer hidden-states prediction. Observing the comparison between BERT and FinBERT, we find these two models perform similarly. 
It shows that pretraining the foundation model on large financial text may not be helpful because the pretraining and fine-tuning stage might use data from different domains.
% It shows that pretraining the foundation model on large financial text may not be helpful because the data used in the pretraining stage and that in the fine-tuning stage might be in different domains.

\paragraph{\textbf{Analysis on \method (Tuning Last)}}
When only fine-tuning the parameters in the last Decoder layer/block on foundation models, we find LLaMA consistently outperforms other three foundation models on \benchmark. The fact that LLaMA-13B is stronger than LLaMA-7B conforms to the findings of LLaMA~\cite{touvron2023llama}. Remarkably, LLaMA-13B with only the last Decoder layer being trainable still has a great risk prediction score, showing the efficacy of \method.

\subsection{Profile Tuning on All Datasets}
\label{subsec:tuning_all}

\begin{figure}[ht]
  \centering
  \includegraphics[width=0.8\linewidth]{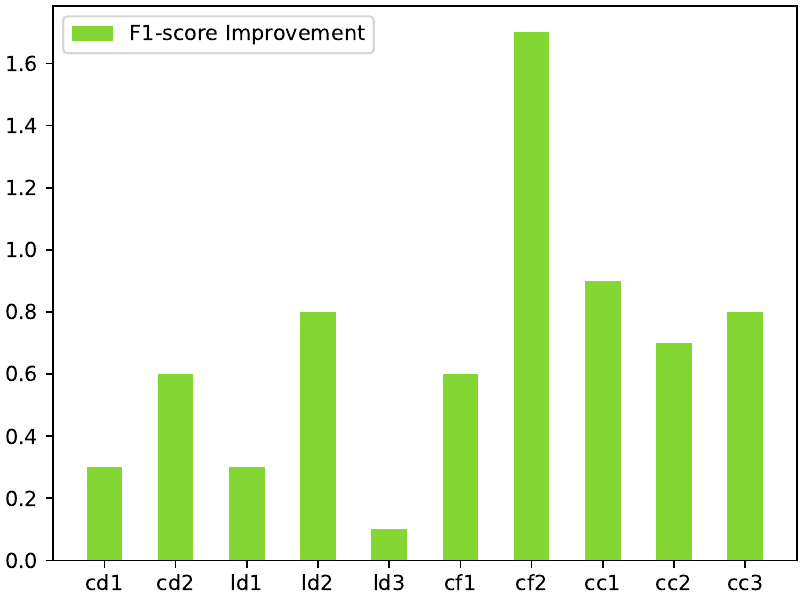}
  % \caption{\textbf{The improvement of \method (Flan-T5-Base) trained on all datasets together over separate datasets.} We report the F1-score gains of training on all datasets in \benchmark instead of training on each separate dataset.}
  \caption{\textbf{The F1-score improvement of \method (Flan-T5-Base) trained on all datasets together over separate datasets.}}
  \label{fig:finpt_tune_all}
  \Description{The overview of \method.}
  \vspace{-5pt}
\end{figure}

Since we transform the tabular data in different datasets into unified text profiles, it is possible to conduct Profile Tuning on all datasets and then evaluate the prediction performance on each test set. As shown in Figure \ref{fig:finpt_tune_all}, we conduct this experiment with the best model in Table \ref{tab:exp_res_main}, i.e., \method on Flan-T5-Base. We can observe that training on all datasets in \benchmark instead of training on each separate dataset brings consistent enhancement. The performance on Dataset \texttt{cf2} benefits the most due to its original low score. F1-score on Dataset \texttt{ld3} improves the least mainly because \texttt{ld3} occupies the majority of the benchmark.

\subsection{Different Strategies}
\label{subsec:tuning_compare}

As mentioned in Section \ref{sec:related_work}, in-context learning and instruction tuning have been a popular trend in utilizing large language models.

For in-context learning (ICL), we input five examples to GPT-2~\cite{gpt2}, Flan-T5~\cite{chung2022flan_t5}, and LLaMA~\cite{touvron2023llama}, hoping that the model generate the correct answer.
% , i.e., ``Yes'' or ``No''
Each example is structured as \textcolor{blue}{``Profile: $\vp_i$ Answer: $\vy_i$''} ($i \in \{1, 2, 3, 4, 5\}$). The last prompt is \textcolor{blue}{``Profile: $\vp_t$ Option 1: Yes. Option 2: No. Answer:''}, where $\vp_t$ is the target profile and the model ought to output the label $\vy_t$ in text (``Yes'' or ``No'').

For instruction tuning (IT), we use an instruction to help the model make predictions: \textcolor{blue}{``Predict whether the following customer is financially risky. $\vp_t$ Option 1: Yes. Option 2: No. Answer:''} Besides, we have tried other informative instruction prompts.

Since the he foundation model often hallucinates and avoids making decisive predictions, we deem the output of ICL and IT to be correct if the right label text (``Yes'' or ``No'') appears in the text. Although it is a relaxed restriction, neither of these two strategies performs well ($<10$ F1-score) compared with other classification baselines.
Nonetheless, we still find the output useful since it provides an additional explanation to help humans analyze the results, as the following example shows:

\textcolor{gray}{``It is difficult to predict with certainty whether a customer is financially risky based on limited information. However, based on the given information, it seems that this customer may not be financially risky. She has a high balance and a good income, and has been a customer for a decent amount of time with a transaction status of 1.0. Additionally, her credit type is average, which is not great but also not poor. Therefore, Option 2: No, she is not financially risky, seems to be a more likely answer.''}

%%%%%%%%%%%%%%%%%%%% Section %%%%%%%%%%%%%%%%%%%%
\section{Conclusion}
\label{sec:conclusion}

In this work, we propose \method, a novel approach for converting tabular financial data into customer profiles, which are subsequently used for predictions after Profile Tuning of large foundation models. In addition, we present \benchmark, a benchmark for financial risk prediction that includes high-quality datasets on three commonly encountered financial risks: default, fraud, and churn. The effectiveness of \method is demonstrated by evaluating a range of strong baseline models on \benchmark. Furthermore, the analytical investigations provide further insights into the application of large foundation models for financial risk prediction.

% \section{Acknowledgments}
% Acknowledgments...

% \section{Appendices}
% Appendices...

% \begin{acks}
% \end{acks}

\bibliographystyle{ACM-Reference-Format}
\bibliography{finpt-sigconf}

% \appendix









\end{document}